# Electrical transport and glassy response in strained thin La$_{0.7}$Ca$_{0.3}$MnO$_3$ films

V. Markovich,[1] E. S. Vlakhov,[2] Y. Yuzhelevskii,[1] B. Blagoev,[3] K.A. Nenkov[4,5]

and G. Gorodetsky[1]

[1]*Department of Physics, Ben-Gurion University of the Negev, P.O. BOX 653, 84105 Beer-Sheva, Israel*

[2]*Institute of Solid State Physics, Bulgarian Academy of Sciences, 72 Tzarigradsko Chaussee Blvd, 1784 Sofia, Bulgaria*

[3]*Institute of Electronics, Bulgarian Academy of Sciences, 72 Tzarigradsko Chaussee Blvd, 1784 Sofia, Bulgaria*

[4]*Institut für Festkörper- und Werkstofforschung Dresden, P.O.Box 270016, O1171, Dresden, Germany*

[5]*International Laboratory for high Magnetic fields and Low Temperatures, 95 Gajowicka Str., 53 421 Wroclaw, Poland*

## Abstract

Magnetotransport properties of La$_{0.7}$Ca$_{0.3}$MnO$_3$ thin films deposited on (100) LaAlO$_3$ substrate were investigated. The balance between the charge ordered insulating phase and ferromagnetic metallic phase may account for a number of glassy features such as: significant hysteresis, memory effects and long-time resistivity relaxation. It was found that the resistance of La$_{0.7}$Ca$_{0.3}$MnO$_3$ thin film decreases significantly upon applying electric current, in a wide temperature range 10 - 220 K. The magnetotransport properties of the strained thin films are discussed in the context of cross-coupling of charge, spin and strain.

Corresponding author: Vladimir Markovich.
Department of Physics, Ben-Gurion University of the Negev, P.O.Box 653, 84105 Beer-Sheva, Israel
Telephone: +972-8-6472456.
Fax: +972-8-6472903;  E-mail: markoviv@bgumail.bgu.ac.il

# I. INTRODUCTION

Perovskite transition-metal oxides exhibit a wide variety of attractive physical properties which originate from mutual cross-coupling among spin, charge and lattice degrees of freedom.[1] The $La_{1-x}Ca_xMnO_3$ (LCMO)-type perovskite-manganese oxides belong to a group of strongly correlated electron systems and exhibit, at the doping range $x = 0.15 \div 0.5$ a metal–insulator (M-I) transition. The above properties result in a colossal magnetoresistance (CMR) behavior. It is known that at a critical doping $x = x_C = 0.225$ of LCMO the ferromagnetic (FM) insulating and orbitally ordered ground state separates from the FM metallic ground state region. In some manganites and in particular in bulk $La_{0.7}Ca_{0.3}MnO_3$, the ferromagnetic transition temperature $T_C$ reaches a value close to room temperature and therefore, thin films of these compounds have been extensively studied due to their potential application as magneto-electronic devices.[2-11] In fact, their magnetic and transport properties at the interfaces may play a crucial role. Various models deal with the modification of the electronic bandwidth due to Mn–O–Mn bond distortion, and crystallographic structure distortion by the interface dead layer. All these aspects have been considered in order to explain the magnetic and transport behavior of thin films compared to the bulk.[2-5,7,11] Charge ordered insulating (COI) regions appear in strained $La_{0.7}Ca_{0.3}MnO_3$ thin films and result in a coexistence of ferromagnetic metallic (FMM) and COI phases, not observed in the bulk.[2-4,9] It is well known that the charge-ordering gap in COI phase collapses upon application of various external perturbations, like magnetic field, electric field/current or high pressure.[12-14] The application of magnetic field and electric current/field to charge ordered systems may result in melting of COI phase and nonlinear transport accompanied with hysteresis and switching phenomena. The properties of thin manganite films depend strongly on their thickness, namely the resistivity increases with decreasing thickness due to the presence of an insulating dead layer at substrate/film interface. The thickness of this dead layer is affected by the substrate, e.g. it is $6.7 \pm 2$ nm for $SrTiO_3$ and $15.3 \pm 4$ nm for $LaAlO_3$ substrates.[4] The induced strains also affect the magnetic anisotropy and the Curie temperature of the manganite thin films. [2-5,15]

Very thin manganite films are of special interest, since they show a number of specific features: non-linear conductivity, strain induced hysteretic behavior, metastable states, and long-time scale dynamics of resistance and memory effects. [2,3,5,6,16,17] Recently[16,17], a pronounced glassy response of the resistance was observed in very thin phase separated $La_{0.8}Ca_{0.2}MnO_3$ film. It was found that applied magnetic and gated electrostatic fields enhances the growth one phase with respect to the other and the resistance varies with the motion of domain boundaries separating the coexisting phases. A similar behavior of resistivity was observed by us (Ref. 18) in low–doped LCMO manganite crystals, in which the competition between magnetic and orbital ordering of coexisting phases results in the appearance of metastable states manifested by metastable conductance, relaxation effects, characteristic two-level noise and resistance memory effects.[18]

Here we report on magnetotransport, electric current effects and relaxation of resistivity in epitaxial thin layers of $La_{0.7}Ca_{0.3}MnO_3$. Though such films were extensively studied in the past the purpose of this paper is to shed light on some aspects concerning the metastable states of x = 0.3, Ca doped thin films. We present various transport measurements carried out in very thin films (150-Å) of $La_{0.7}Ca_{0.3}MnO_3$, grown on $LaAlO_3$ (LAO) (100) substrate. The correlation between the transport properties and magnetic characteristic of these thin films is also discussed. This study was motivated by previous investigations[2,3,9,10] of $La_{0.7}Ca_{0.3}MnO_3$ films of 150-Å thickness, in which a switching from semiconducting to metallic behavior and hysteresis phenomena were observed in high magnetic fields. The present transport experiments were performed under various measurement protocols, thereby providing additional information on the transport mechanism in these very thin films. Within the coarse of these investigations we also study the effect of current-induced resistivity changes.

## II. EXPERIMENTAL

Thin films of $La_{0.7}Ca_{0.3}MnO_3$ were grown on (100)-oriented $LaAlO_3$ (LAO) substrate using magnetron sputtering. More details of the deposition, x-ray diffraction, magnetization and magnetotransport studies are described elsewhere. [7,9,10] The analysis of the x-ray diffraction show that the mismatch of -1.81%, leads to a compressive strain in plane and tensile strain in the off-plane direction. In order to investigate the structure of the deposited 150-Å/LAO $La_{0.7}Ca_{0.3}MnO_3$ films we have elaborated the deposition regime as well as the growth rate because the layer thickness and oxygen deficiency may crucially influence the magneto-transport properties of manganite layers. The layers were characterized by x-ray diffraction and Rutherford backscattering. It was found that the deposited layers are exhibit reproducible magneto-transport properties, during years. The magnetotransport properties were measured in using a four-probe configuration at temperatures 10 – 300 K and magnetic fields up to 15 kOe applied along the current direction. The preparation procedure and the experimental setup for simultaneous measurements of resistance and current-voltage characteristics were previously described.[18] In our experiments, evaporated gold pads for wiring with a separation of about 0.3 mm between the voltage ($V$) pads were used. In order to avoid Joule heating all of the resistivity measurements, excluding the measurements of the resistance as a function of current, were recorded with a relatively low current of 1 µA. Isothermal measurements were carried out by the following procedure: the samples were cooled down from room temperature at zero field (ZFC) or cooled under applied fields (FC) to $T \sim 10$ K, then heated to the desired temperature of the measurements.

## III. RESULTS AND DISCUSSION

Figure 1 shows the temperature dependences of dc resistivity ($I = 1$ µA) of the $La_{0.7}Ca_{0.3}MnO_3$ thin film (150-Å) upon slow cooling and subsequent heating, under various magnetic fields. The behavior of the resistivity in zero magnetic fields (ZFC protocol) appears to be insulating in the whole temperature range. Previous measurements of these thin films have shown that metallic behavior ($dR/dT > 0$) appears at $T < 220$ K under an applied field of 75 kOe.[9] An application of modest magnetic fields result only in a small decrease of resistivity at low temperatures. One should note that the resistivity at low temperatures exhibits a complex hysteresis between curves recorded upon cooling and subsequent heating and also high level of noise at temperatures below 70 K. A pronounced hysteresis was observed in wide temperature range also at zero magnetic field. With increasing fields the form of hysteresis remarkably changes, see Fig. 1. The inset in Fig. 1 presents the resistivity recorded under the following protocol: after a slow cooling to $T = 10$ K in magnetic field $H = 14.5$ kOe, the magnetic field was turned off and the resistance was measured in zero magnetic field at a slow heating rate (~0.5 K/min). It was found that at temperatures 10 K - 35 K the resistivity changes only slightly, and then increases with increasing temperature (35 - 80 K), while exhibiting a noisy resistance behavior. Such a behavior may indicate that the sample resistance undergoes irreversible changes, a behavior, which resembles thermo-remnant magnetization in spin-glasses.[19,20] It should be noted that no hysteresis effects were observed above 200 K, see Fig. 1. One may conclude from the above results that at zero field and at temperatures 10 -300 K the film resides in an insulating state, in compliance with previous measurements [2,9,10] An application of magnetic field converts some volume fraction to a metallic state, therefore resulting in a decrease of the resistivity. One should note that the value of the resistivity and the memory effects depend also on the cooling protocol.

Figure 2 shows the resistance vs. magnetic field at various temperatures after ZFC to 10 K and subsequent warming to the desired temperature at no field. It was found that the resistance reaches a lower resistivity state after applying of a magnetic field; see Fig. 2(a). In our case the field driven resistivity changes exhibit a pronounced hysteresis mostly due to relaxation in the electronic transport, which will be discussed later. The effect of negative magnetoresistance becomes more pronounced with decreasing temperatures and increasing magnetic fields, e.g. application of magnetic field of 75 kOe at $T$ = 50.5 K results [9,10] in a very sharp transition from an insulating state to a metallic one, exhibiting a drop of four orders of magnitude. This may be also an indication that the volume of the FM phase at $H$ = 75 kOe exceeds the percolation thereshold ~17%. Based on the observed values of the saturation magnetization of 150-Å $La_{0.67}Ca_{0.33}MnO_3$/LAO at 5 K (~ 1.8 $\mu_B$/f.u., which is about of 50% of the expected $M_{sat}$/f.u. = 3.67 $\mu_B$) Biswas *et al* [2] have concluded that about a half of the volume of the manganite layer is not FM at low temperatures. An almost linear dependence of the resistance vs. magnetic field was observed at $T$ = 150 K, see Fig. 2(b) and hysteresis effects are probably masked by the relatively strong effect of the magnetic field. The magnetoresistance (MR= $((R(0)-R(H)/R(H))$ at 150 K reaches a value of 86% in a magnetic field of 14.5 kOe see Fig. 2(b). At higher temperatures the effect of magnetic field decreases though, hysteresis and relaxation processes become more visible, see Fig. 2(c). Three successive sweepings of magnetic fields show progressive small decrease of resistance, resembling some kind of the training phenomena in magnetic fields.[21] Melting and disappearing of COI state in a similar thin film of 150-Å $La_{0.67}Ca_{0.33}MnO_3$ deposited on $LaAlO_3$ (LAO) were previously observed [2] at low temperatures and at high magnetic fields, resulting in metastable states after the magnetic field was removed. Created by this means a metastable state consisted FM clusters and antiferromagnetic charge-ordered matrix was observed, resulting in a slow relaxation dynamics of the resistivity at temperatures below 70 K. [2] At low temperatures small FM clusters are isolated and frozen in different directions and do not form percolating network in modest magnetic field. Only under an application of high enough magnetic field of ~ 75 kOe [9,10] a percolation among disconnected clusters occurs. Namely, at $T$ < 50 K, and after removing the magnetic field $La_{0.7}Ca_{0.3}MnO_3$ thin film sample resides in a "frozen" cluster glass-like state that does not relax. At 50 < $T$ < 80 K our sample exhibit a short time relaxation with extremely high level of noise, see inset in Fig. 1 and Ref. 16. At higher temperatures (90 < $T$ < 130 K) and after turning on and off of magnetic fields the short time relaxation is replaced by a long time one, see Fig. 3. We have fitted the time dependence of the resistivity after turning on and out with a stretched exponential form $R(t) = R_0 - R_1 \exp[-(t/\tau)]^\beta$. As pointed out by Biswas *et al.* [2] such long-time relaxation of the resistance has been observed in thin films and bulk charge-ordered manganite systems, but the exact physical significance of the parameters $\tau$ and $\beta$ is not fully understood. It was suggested by above authors that metastable states separated by energy barriers with a wide distribution of energies exist in the bulk and in thin layers. This leads to a distribution of relaxation time and the stretched exponential behavior of the relaxation. An estimation of $\beta$ at various temperatures yields a maximal value of 0.45 – 0.5 at the temperature range 50 - 80 K. These values are found to be in rough agreement with the results of Ref. 2. Nevertheless in distinct contrast with Ref. 2, the long-time relaxation resistance was observed here in a wider temperature range of 50 - 200 K, see Figs. 1- 3.

In this study we also searched for nonlinear resistivity features, which often have been observed in phase separated thin films,[12,22-24] tunnel junctions,[25] polycrystalline samples[14] and single crystals of manganites.[13,18,26] A difficult problem one unavoidably faces upon studying of the effect of electric current/field on the resistivity is Joule overheating, especially for semiconducting-like ($dR/dT < 0$) behavior. In such a case the current-induced overheating may cause a significant heating of the sample and therefore nonlinear $V$ - $I$ characteristics.[24,27,28] Short pulse current measurements may sometimes be appropriate to reduce effects of Joule heating in systems that do not exhibit long time relaxation. This is not the case studied here. Therefore, only DC measurements were carried out on our sample. The resistance of the film under a magnetic field of 14.5 kOe is considerably reduced and depends only slightly on temperature in wide temperature range 10 - 180 K, see Fig. 1. In all of our measurements the current does not exceed 40 µA (Figs. 4 and 5). The temperature increase of the sample at temperature $T$ can be evaluated by the expression: $\Delta T(T,I) = (2I^2\rho(T + \Delta T)/Sk_{sub}(T + \Delta T)$, where $S$ – is the cross section of the thin layer and $k_{sub}$ is the thermal conductivity of the substrate. [24,27] Using the experimental values of the resistivity and the average value of $k_{sub}$ = 0.15 W cm$^{-1}$ K$^{-1}$ in the temperature range 10 -170 K, [24,27] one finds out that the increase of temperature for a current of 40 µA does not exceed ~ 0.2 K. This evaluation agrees well with the calculation of the temperature increase for $Pr_{0.5}Ca_{0.5}MnO_3$ thin film (800-Å) on LAO substrate, [24] indicating that only a current of few hundreds µA may result in considerable overheating. It should be noted that in above equation the cross section of the whole film was taken into account, whereas in the case of phase-separated films the actual cross section through which the current flow corresponds only to the metallic phase, and as a result, the effective cross section is reduced at low temperatures. Therefore, it seems on the first glance that our evaluation of the temperature increase is rather underestimated since local overheating may possibly be considerably higher. On the other hand, in the case of thin films most of the temperature gradient is in the substrate [24,27,28] and this factor affects also the process of thermal dissipation. In addition to above arguments, it should noted that the measurements of $R$ vs. bias current and $V$- $I$ curves were carried out at presence of magnetic field $H$ =14.5 kOe [see Figs. 4(a) and 5(a,b)], while the resistance varies relatively slightly over wide temperature range, 10 - 180 K. This situation is likely indicative of almost temperature independent proportion of metallic regions in wide temperature range when magnetic field $H$ =14.5 kOe is applied. Thus, in these conditions the effective Joule heating may only slightly depend on the temperature in above temperature range. Isothermal resistance measurements show that electronic current causes to a significant decrease of the resistivity at the whole temperature range 10 - 250 K, this effect becomes more pronounced with decreasing temperature. The temperature variation of the relative change in the resistance at two definite currents (20 µA and 40 µA) exhibits a similar behavior. It also appears that the curves presented in Fig. 4(b) resemble very closely the temperature dependence of FC magnetization of our films. This fact is an indication, that magnetotransport properties of $La_{0.7}Ca_{0.3}MnO_3$ thin film are governed by the magnetic state of the film. One may also suggest that the magnetization and magnetotransport are similarly affected by the FM clusters and strain.

The FC magnetization starts to increase below $T \sim 250$ K (Fig. 4(b), signaling an onset of ferromagnetic transition, but this rise is much slower than that observed in thicker films.[3,5,8] Generally, the reduction in the film thickness results in a progressive decrease of the Curie temperature to lower values, where the Curie temperature $T_C$ is defined by the temperature of the maximum slope of the magnetization, here $T_C$ is of about 180 K. Such behavior of magnetization is alike with the previous one observed for $La_{0.7}Ca_{0.3}MnO_3$ of similar thickness deposited on LAO substrate. [2,3,5,8] This may be an indication for a smeared out ferromagnetic transition caused by a wide distribution of Curie temperatures in different FM clusters, which leads to inhomogeneous electronic and magnetic states below $T \sim 180$ K. The ZFC – FC magnetization appears to be irreversible below $T_{ir} \sim 180$ K [(see Fig. 4.(b)]. This may be attributed to the formation of spin/cluster glass-like state. It is also interesting to point out that FC magnetization continues to increase below the irreversibility temperature $T_{ir}$, at which both ZFC and FC curves merge, exhibiting a characteristic feature of cluster glass systems.[29,30,31]

Figures 5(a,b) show the voltage-current $V$ - $I$ characteristics recorded at various temperatures after FC cooling down to 10 K and subsequent heating to the desired temperature under a magnetic field of $H = 14.5$ kOe. It is obviously seen that the $V – I$ characteristics is almost linear at $T = 250$ K, and become progressively nonlinear with decreasing temperatures, see Fig. 5(b). The nonlinearity observed in the $V – I$ curves may signify the presence of tunneling mechanism, described in particular by the Glazman – Matveev (GM) theory.[32] According to the GM theory [32] for $eV \gg k_BT$ the $V – I$ dependence of a thin amorphous film is expressed by

$$I = (G_0 + G_1)V + \sum_{n=2}^{\infty} a_n V^{n+1-2/(n+1)} \qquad (1)$$

where first term represents the direct and resonant tunneling via one impurity, while the nonlinear $a_n$ terms describe an inelastic multi-step tunneling via localized states; the power index for n =2 is 7/3 and for n =3 is 7/2. The experimental $V - I$ dependences at various temperatures were fitted to the following expression: $I = G_0 V + G_1 V^\alpha + G_2 V^\beta$,[33] with the power indexes: $\alpha$ = 7/3 and $\beta$ = 7/2, see Fig. 5(c). The fitting for three temperatures is presented in Fig. 5(c), showing a fairly well fit at 10, 100 and 150 K. It appears that GM model may be applied to present study of the 150-Å $La_{0.67}Ca_{0.33}MnO_3$/LAO layer. This observation is in line with a previous analysis of nonlinear conduction in thin film structures using the GM model.[32] As temperature rises above 150 K, first linear term progressively increases with increasing temperature and at temperatures above $T \sim 250$ K it dominates the transport mechanism.

It has to be emphasized that not like in the case of artificial tunnel junctions, one cannot provide an absolute proof for a tunneling process in phase separated materials. Indirect evidence can be obtained from the observations of nonlinear voltage-current ($V - I$) characteristics and their temperature evolution. These will have to be matched with appropriate tunneling model, though in our case the presence of spin/cluster glass-like features may also affect the behavior of tunneling characteristics at low temperatures. The intrinsic tunnel barriers in manganites are associated with extended crystalline defects such as grain or twin boundaries. They may exhibit pronounced effects when percolating metallic paths become separated by insulating regions of the less conducting phase. The appearance of an insulating regions may create variation in the bond angle and the double exchange mechanism.[34] It may lead also to electronic band bending due to strain fields associated with such defects,[35] or to phase separation at the internal interface.[36] It has been proposed that in thin $La_{0.67}Ca_{0.33}MnO_3$ film the shape and in plane alignment of FM and AFM clusters may be affected by granularity and by twin structure in the volume of film during growth.[37] As was shown by Biswas et al.[3] thin films (150-Å) of La$_{0.67}$Ca$_{0.33}$MnO$_3$, grown on LAO substrate has an island growth mode with highly nonuniform distribution of the strain. Such variation of strain in the film may also result in the migration of the constituent atoms, leading to compositional inhomogeneity.[3,5] It has been suggested that the top of the islands are under low strain and these regions are ferromagnetic and metallic, whereas the large strain at the edge of the islands results in insulating and possibly charge-ordered state. This leads us to suggest the possibility of the development of various FM phases at the top of islands, which act as conducting banks for intrinsic tunnel junctions, whereas localized levels may be located in insulating edge of the islands and at migrated ions inside the tunneling barriers. A similar scenario of tunneling through localized levels in the interglanular barrier was proposed in theoretical works[38] and was exploited for interpretation of transport properties of La$_{1-x}$Ca$_x$MnO$_3$ (x = 0.3,[39] 0.33[37]) and La$_{0.6}$Sr$_{0.4}$MnO$_3$[40] films.

According to the phase diagram[2] of La$_{0.67}$Ca$_{0.33}$MnO$_3$ 150-Å thin layer, it comprises below $T_{CO}$ ~ 150 K ($H$ = 0) low strain FM regions and high strain CO regions. The two coexisting phases of the film are coupled via localized structural distortions, manifested by slow relaxation dynamics of the resistance. The evolving of the resistance upon turning magnetic fields on and off appears to be logarithmical with time. Such a behavior, observed recently in various manganite systems,[18,41,42] is typical for cluster/spin glasses. It is believed that an application of magnetic field changes the ratio of the coexisting phases. However, at modest magnetic field and at low temperatures the resistance is practically temperature independent and the absence of the resistance relaxation resembles a "frozen" spin glass-like state below freezing temperature.[18,42]

In order to provide an insight into the variation of ferromagnetic fraction $f$ under applied current/voltage as an alternative to GM approach, the general effective medium (GEM) was used.[43,44] Recently, there have been several attempts at analysis of transport properties of CMR manganites as binary metal–insulator mixtures.[33,45-47] For electric ($\sigma_E$) or thermal ($k_E$) conductivity of a binary M-I mixture McLachlan [43] proposed the GEM equation,

$$f \frac{(\sigma_M^{1/t} - \sigma_E^{1/t})}{\sigma_M^{1/t} + A\sigma_E^{1/t}} + (1-f) \frac{(\sigma_I^{1/t} - \sigma_E^{1/t})}{\sigma_I^{1/t} + A\sigma_E^{1/t}} = 0 \qquad (2)$$

where $f$ is a volume fraction of a ferromagnetic metallic phase, $\sigma_M$ and $\sigma_I$ are the conductivities of the metallic and insulating phases, respectively. The effective conductivity, measured experimentally, is defined as $\sigma_E$, $t$ is a critical exponent. $A = (1 - f_C)/f_C$, where $f_C$ is the percolation threshold. The value of the exponent $t$ in the above equation ranges, for a three dimensional lattices between 1.65 and 2.0, as demanded by universality. [44] It was recently found that the value $t = 2$ describes well various manganite systems: polycrystals,[46,47] single crystals [33] and thin films. [45] For conductivity, we assumed $\sigma_M(T)$, the value of conductivity of relatively thick (150 – 200 nm) $La_{1-x}Ca_xMnO_3$ (x ≈ 0.3) films with metallic behavior at low temperatures.[48] For $\sigma_I(T)$ we assumed the value of the conductivity of our film at zero magnetic field, see Fig. 1. In similarity with results of Wu et al., [45] we found that modeling of equation (2) using $f_C$ = 0.22 – 0.24 yields better fitting to our data. As pointed out in Ref. 45, the difference between this value and the accepted 3D value of 0.16 may be attributed to the quasitwo-dimensional nature of the thin film. On the other hand, some errors also arise from differences between our assumed $\sigma_M$ and $\sigma_I$ and the real values. [45] The calculated $f(V)$ and the resistivity vs. the applied voltage for $T = 50$ K is presented in Fig. 6. It should be emphasized, that $f$ varies relatively slightly with the applied voltage. This behavior may be qualitatively understood as follows: an applied magnetic filed of $H$ =14.5 kOe converts some volume fraction to a metallic state, pushing the film towards percolation threshold and therefore resulting in a significant decrease of the resistivity, see Fig. 1. The effect of electric field/current is relatively weaker in comparison with effect of magnetic field at least in the restricted range of applied current used, see Fig. 6.

Recent analysis of transport properties of $La_{0.67}Pr_xCa_{0.33}MnO_3$ thin films using effective medium theory have shown that these films cannot be described by a simple metal-insulator mixtures. [45] One plausible reason suggested in Ref. 45 is the presence of a third phase intermediating between the two, being ferromagnetic and insulating. A smeared out Curie temperatures (see Fig. 4(b)), may also be an indication for a presence of FM clusters with different magnetic and electronic properties. It appears therefore that the description of system as a textured material with the multiscale, multiphase coexistence (see Ref. 49) seems more reasonable than a simple fixed mixture of conducting and insulating media.

The results, reported above (Fig. 3) are found to be of much similarity to the glassy response of another thin films $La_{0.8}Ca_{0.2}MnO_3$ 82-Å under external magnetic and electric fields. [16,17] The glassy behavior observed in $La_{0.8}Ca_{0.2}MnO_3$ 82-Å films was discussed in terms of metastable energy landscape [49] with hierarchical energy barriers for relieving the strain. Such a model supports a glass like response of the sample to external magnetic field due to partial melting of the charge-ordered phase and the change of the phase boundaries between COI and FM phases. One should note that domain boundaries shift irreversibly through a metastable pinning landscape, yielding conductance changes by the application of magnetic field. [16,17] Pinned domain walls between COI and FM phases can be directly coupled to the strain fields in the film. The observed complex hysteresis in resistivity between warning and cooling (see Fig. 1) suggest that the transition is of a first order. Moreover, such a complex hysteresis, separated into two branches may be attributed to the double phase transitions: charge-ordering state at $T \sim$ 150 K and the transition to "frozen" cluster/glass like state at $T \sim 70$ K. Its occurrence over wide temperature interval is somewhat puzzling and there is no a good explanation for such a feature at this moment. It is likely that the observed hysteretic behavior is driven by a fine balance of competing charge order with strong tendency to localization, FM interactions and strain energy.

Contrary to the behavior of $La_{0.8}Ca_{0.2}MnO_3$ films [16,17] and single crystals,[18] an applied current in $La_{0.7}Ca_{0.3}MnO_3$ films does not produce hysteretic and memory effects. The strongest effects of electric field/current should fall at the boundaries between metallic and insulation regions, where the electric field is maximal. At the boundaries the accumulated charges in metallic regions driven by electric field/current would then undergo a force that would literally pull the boundaries further into insulating regions, and rising the fraction of metallic phase. [16] This shift of the boundaries may occur as reversible process at least in range of relative low currents. Similar scenario was suggested just recently for $La_{0.67}Ca_{0.33}MnO_3$ thin films 1000-Å, deposited on $SrTiO_3$ substrate.[50] It was argued [50] that the current passing through the FM regions is polarized and then injected into the non-FM insulation regions (COI in our case), keeping their polarization within a certain depth and leading to an increase of the FM conduction regions. In principle, both models applied in this article (GM and GEM) are supplementary, supporting the coexisting mechanisms that describe the effect of applied electric current/field on the resistivity, namely inelastic tunneling of charge carriers across localized states in the barriers separating FM regions (GM) and shift of boundaries between metallic and insulating domains towards the insulating regions (GEM).

In conclusion, we have studied the effects of magnetic field and current on the transport properties of $La_{0.7}Ca_{0.3}MnO_3$ thin films deposited on $LaAlO_3$ substrate. We have performed measurements of resistivity using various protocols. The results observed reveal the glassy features observed at the temperature range of phase separation. The metastable states that manifested themselves through the long time resistivity relaxation, switching and memory effects after turning on and off magnetic field were observed. The nonlinear electronic transport behavior of our thin layer is also a manifestation of mixed phase behavior.

**FIGURE CAPTION**

Fig. 1. (color online) Temperature dependence of the resistivity of $La_{0.7}Ca_{0.3}MnO_3$ thin film, measured at cooling and subsequent heating in various magnetic field. Inset shows the temperature dependence of the resistivity recorded at cooling at a magnetic filed of $H = 14.5$ kOe. After turning off the magnetic field at $T = 10$ K the temperature dependence at warning is recorded in zero magnetic field.

Fig. 2. (color online) Resistance vs. magnetic field measured after ZFC to $T = 10$ K and subsequent heating to desired temperature of measurements: (a) $T = 90$ K. Field sweep starts at point 1; (b) $T = 150$ K; (c) $T = 200$ K, three successive runs. The curves are not reversible.

Fig. 3. (color online) Resistance after reversing on and off the magnetic field of 14.5 kOe at $T = 120$ K, after ZFC procedure, see text. Insets show the relaxation of the resistance in extended scale.

Fig. 4. (color online) (a) Resistance as a function of current at various temperatures; (b) Temperature variation of the resistance for $I = 20$ and 40 µA. For comparison the temperature dependence of ZFC and FC magnetization measured in $H = 50$ Oe is also included (Ref. 10).

Fig. 5 (color online) (a,b) Curves of current vs. voltage after FC to $T = 10$ K in a magnetic field of $H = 14.5$ kOe and subsequent heating to the desired temperatures; (c) Experimental points fitted (solid line) by the multi-step tunneling GM theory.

Fig. 6 (color online) Resistivity vs. applied voltage and FM volume fraction obtained for $H = 14.5$ kOe and $T = 50$ K. The dashed line indicates the critical value $f_C$ taken for modeling by equation (2).

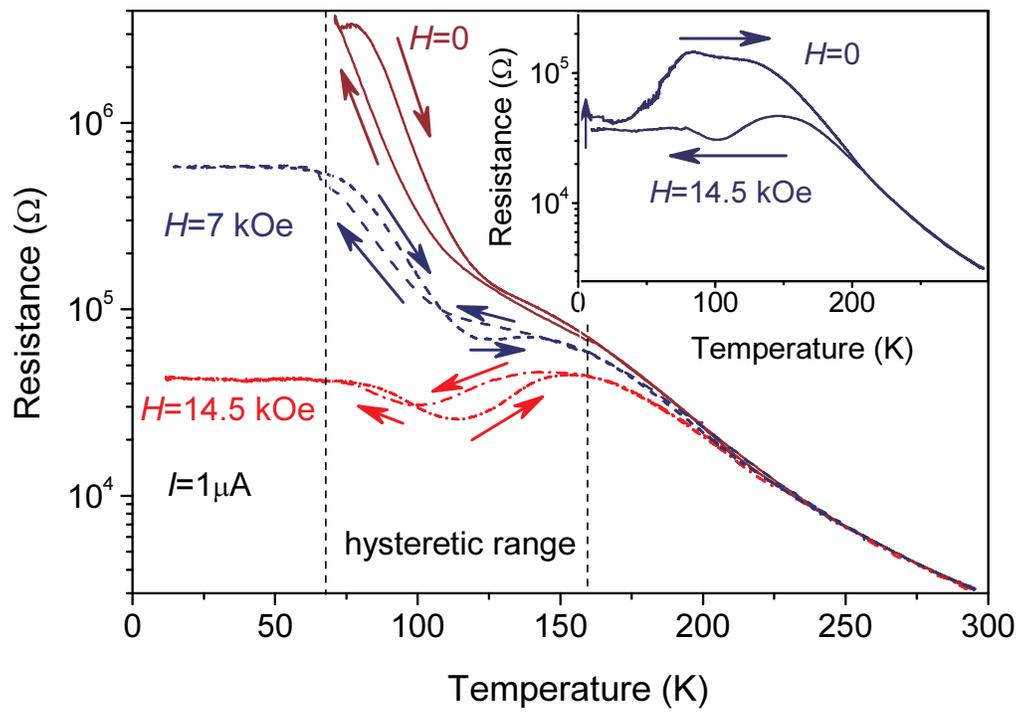

Fig. 1

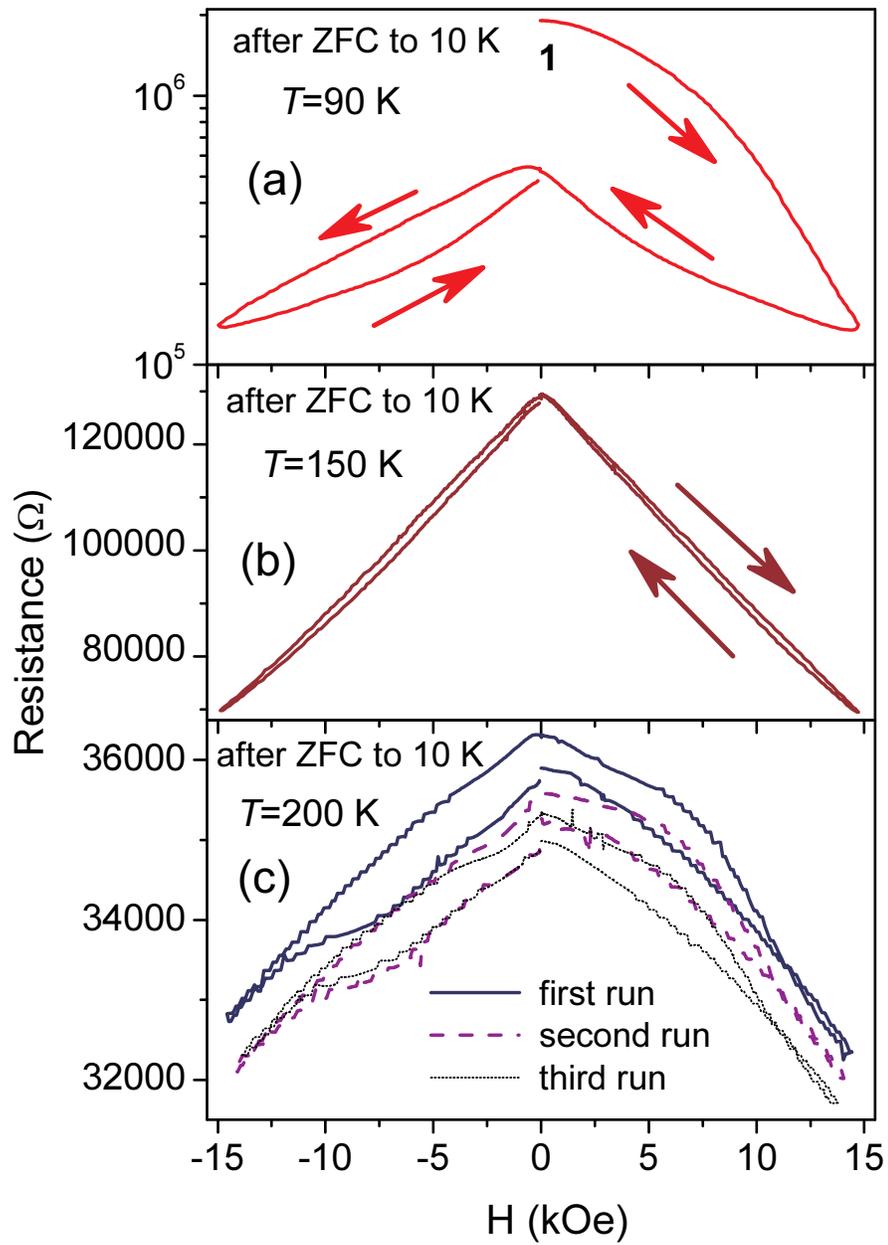

Fig. 2

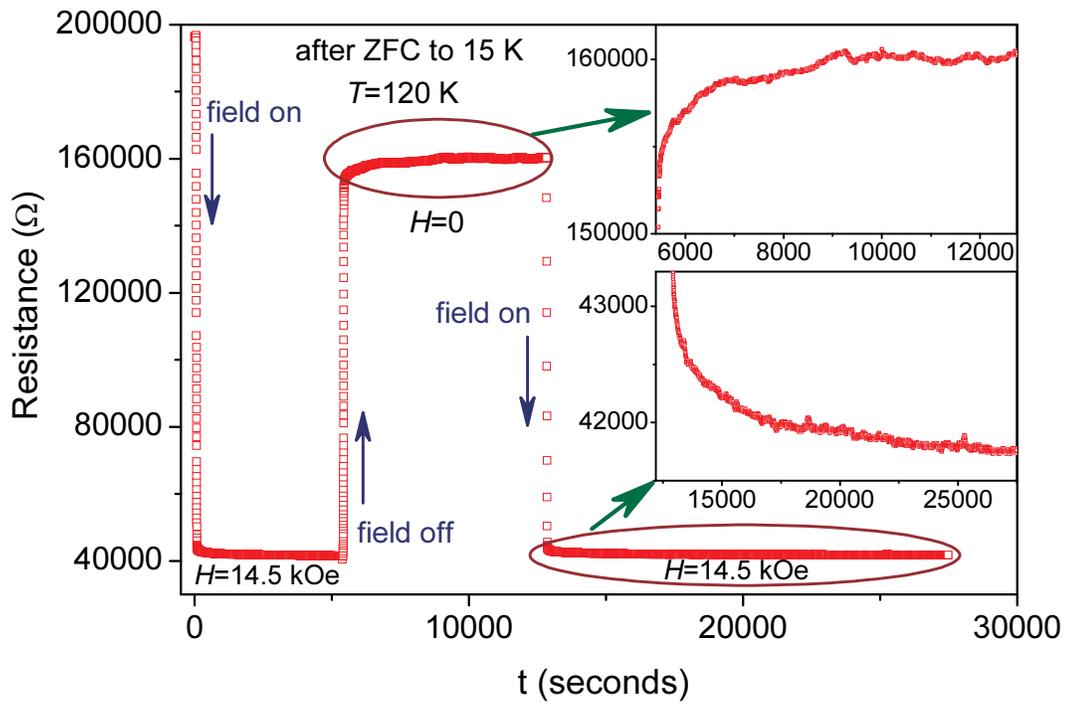

Fig. 3

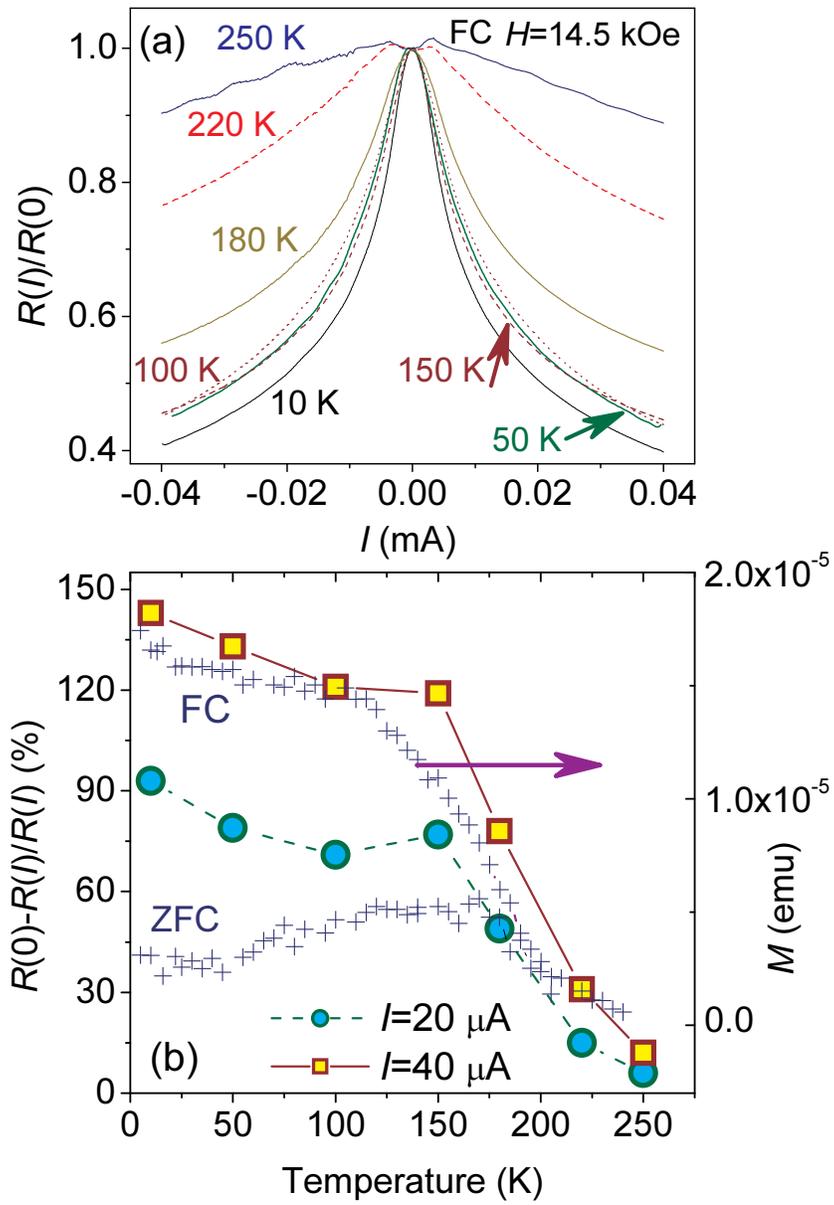

Fig. 4

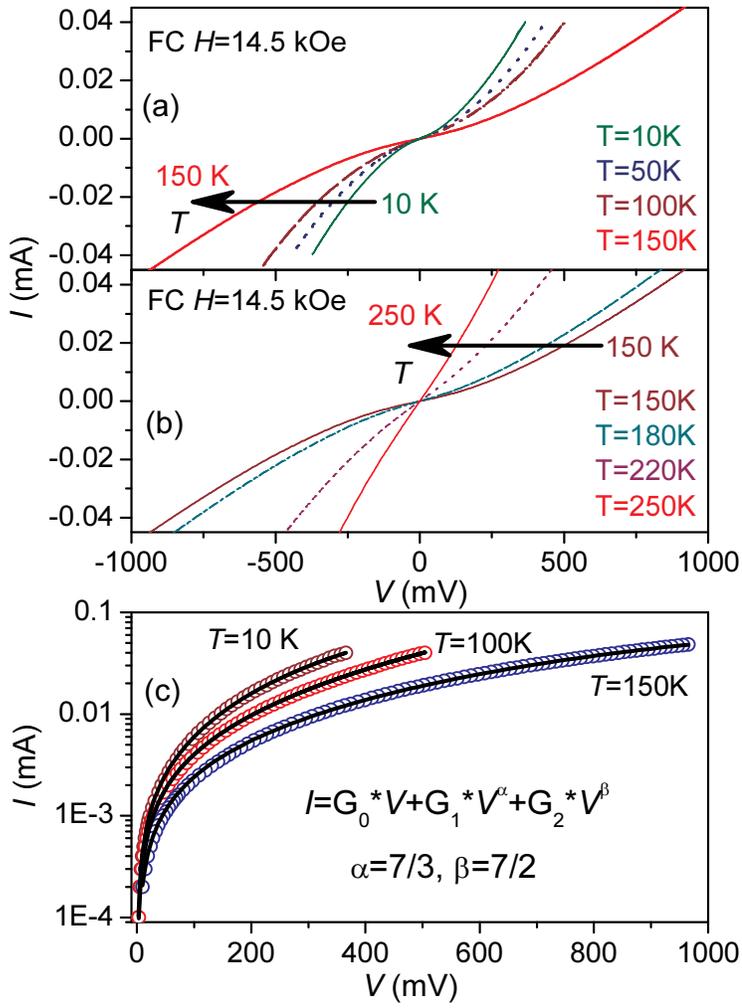

Fig. 5

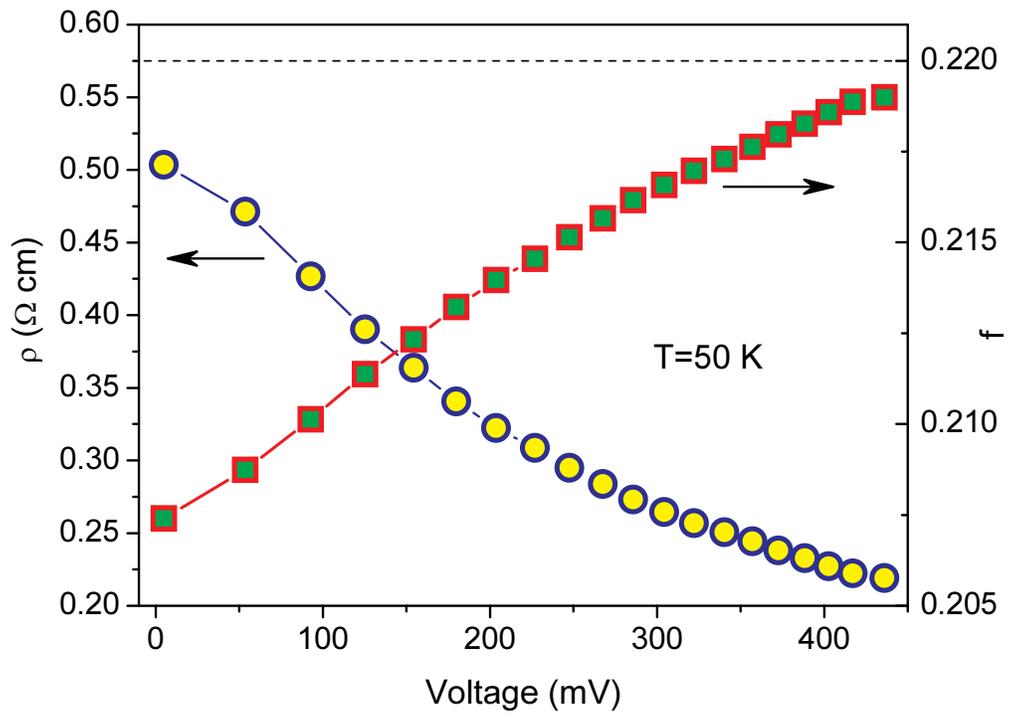

Fig. 6